\documentclass[preprint,showpacs,preprintnumbers,amsmath,amssymb,floatfix]{revtex4}

\usepackage[small]{caption} 
\usepackage{graphicx}
\usepackage{dcolumn}
\usepackage{bm}
\usepackage{url}
\usepackage{amsmath}
\usepackage{verbatim}  
\usepackage{relsize}

\newcommand{\pb}{$^{208}\mathrm{Pb}^{82+}$}

\newcommand{\nb}{N}

\newcommand{\nip}{n_\mathrm{IP}}
\newcommand{\kb}{k_\mathrm{b}}
\newcommand{\frev}{f_\mathrm{rev}}

\newcommand{\exy}{\epsilon_{xy}}

\newcommand{\exyi}{\epsilon_{xyi}}
\newcommand{\exyj}{\epsilon_{xyj}}

\newcommand{\exI}{\epsilon_{x1}}

\newcommand{\eli}{\epsilon_{li}}

\newcommand{\vI}{\vect{v}_1}
\newcommand{\vII}{\vect{v}_2}
\newcommand{\drm}{\mathrm{d}}

\newcommand{\betxy}{\beta_{xy}}
\newcommand{\betSt}{\beta^*_{xy}}
\newcommand{\alphSt}{\alpha^*_{xy}}

\newcommand{\tibsxy}{T_{\mathrm{IBS},xy}}

\newcommand{\tibsl}{T_{\mathrm{IBS},l}}
\newcommand{\tradz}{T_{\mathrm{rad},z}}
\newcommand{\tradxy}{T_{\mathrm{rad},xy}}
\newcommand{\tcoll}{T_{\mathrm{c}}}
\newcommand{\tcollZer}{T_{\mathrm{c0}}}

\newcommand{\rhoRemain}{\rho_\mathrm{rem}}

\newcommand{\xIo}{x_{10}}
\newcommand{\yIo}{y_{10}}
\newcommand{\zIo}{z_{10}}
\newcommand{\xIIo}{x_{20}}
\newcommand{\yIIo}{y_{20}}

\newcommand{\sigtot}{\sigma}

\newcommand{\lum}{\mathcal{L}}
\newcommand{\lumsc}{\mathcal{L}_{sc}}

\newcommand{\vect}[1]{\vec{#1}}

\newcommand{\madx}{\textsc{mad-x}}

\begin{document}


\title{Emittance increase caused by core depletion in collisions}

\author{R.~Bruce}
\email{roderik.bruce@cern.ch}

\date{\today}

\begin{abstract}
A new effect is presented, which changes the emittance during colliding-beam operation in circular colliders. If the initial transverse distribution is Gaussian, the collision probability is much higher for particles in the core of the beam than in the tails. When small-amplitude particles are removed, the remaining ones therefore have a larger transverse emittance. This effect, called core depletion, may cause a decrease in luminosity. An approximate analytic model is developed to study the effect and benchmarked against a multiparticle tracking simulation. Finally, the time evolution of the intensity and emittances of a \pb bunch in the Large Hadron Collider (LHC) at CERN is calculated, taking into account also other processes than collisions. The results show that integrated luminosity drops by 3--4\% if core depletion is taken into account. It is also found that core depletion causes the transverse emittance to be larger when more experiments are active. This observation could be checked against experimental data once the LHC is operational.

\end{abstract}


\maketitle

\section{Introduction}
During operation of a circular collider, such as the Large Hadron Collider (LHC) at CERN~\cite{lhcdesignV1}, particles are continuously removed or redistributed within the beams by a number of different processes, e.g.  the collisions, intrabeam scattering (IBS), radiation damping and scattering on rest gas.  The coupled effect of these processes determine the time evolution of the bunch intensities, emittances and the luminosity during a store.

In this text, the effect of the collisions on the beam distribution is studied using numerical parameters for \pb operation in the LHC. These parameters are given in Table~\ref{tab:lhc}. If the initial transverse bunch distribution is Gaussian, the interaction probability is higher in the centre of the bunch. Therefore, the ratio of the number of particles removed in a bunch crossing to the initial number of particles over some small transverse distance is much higher in the central part of the bunch than in the tails. This leads to a depletion of the core of the beam. The transverse emittance of the remaining particles is therefore increasing in the absence of damping, which in turn leads to a decreasing luminosity. In principle a similar effect is present in the longitudinal plane, since the hourglass effect decreases the collision probability for particles ahead of or behind the synchronous particle. This effect is however extremely small and can be safely neglected under most realistic machine conditions.

\begin{table}[tbh!]
\begin{center}
\label{tab:lhc}
\caption{ Nominal parameters for \pb ion operation in the LHC, given for the beginning of store taken from Ref.~\cite{lhcdesignV1}. The luminosity reduction factors have been calculated using Eq.~(\ref{eq:rtot-gauss}). Three interaction points are considered for ion collisions: \textsc{alice} (IP2), \textsc{atlas} (IP1) and \textsc{cms} (IP5).}
\begin{tabular}{|l|c|}
\hline
Parameter               & Value  \\
\hline \hline
Ion species             & \pb \\ \hline
Beam energy             & 2759 GeV/nucleon  \\ \hline
Lorentz factor $\gamma_\mathrm{rel}$   & 2963.5 \\ \hline
Bunch intensity $N_b$   & $7\times10^7$ \\ \hline
Bunches per beam        &  592  \\ \hline
Normalized transverse rms emittance & 1.5 $\mu$m \\ \hline
Long. emittance at 4 $\sigma$    &  2.5 eV s/charge \\ \hline  
rms bunch length        & 7.94 cm  \\ \hline
rms energy spread       & $1.1\times 10^{-4}$ \\ \hline
N.o. active interaction points (IPs) & 1--3 \\ \hline
Total interaction cross section $\sigtot$ & 515 b \\ \hline
Optical function $\betxy^*$ at IP2    & 0.5 m \\ \hline
Optical function $\betxy^*$ at IP1 and IP5 & 0.5 m \\ \hline
Crossing angle at IP2    &  70 $\mu$rad  \\ \hline
Crossing angle at IP1 and IP5    &  285 $\mu$rad  \\ \hline
Geometric luminosity reduction $R$, IP2 & 0.974 \\ \hline
Geometric luminosity reduction $R$, IP1 and IP5 & 0.825 \\ \hline
Peak luminosity         & $10^{27}$ cm$^{-2}$s$^{-1}$ \\ \hline
RF harmonic number $h$ & 35640 \\ \hline
RF gap voltage         & 16 MV   \\
\hline
\end{tabular}
\end{center}
\end{table}

To study the transverse core depletion effect in detail, an analytic model is developed in Sec.~\ref{sec:analytic_model} under the assumption that the distribution remains approximately Gaussian but changes in size over time. To test the validity of this and other assumptions in the model, it is compared with a multiparticle tracking code is described in  Sec.~\ref{sec:tracking}. This code includes only collisions, synchrotron motion and betatron motion, neglecting other effects such as IBS. This somewhat artificial situation is analyzed in order to see the isolated effect of the core depletion.

The analytic model is compared with the tracking in Sec.~\ref{sec:comparison} and a very good agreement is found using numerical parameters for \pb operation in the LHC. 
In Sec.~\ref{sec:ODEs}, finally, the analytic model is extended to include other processes changing the beam distribution. The coupled behaviour results in a system of ordinary differential equations (ODEs), which has to be solved numerically in analogy with Ref.~\cite{epac2004}. Solutions with and without core depletion are presented.

\section{Emittance increase from collisions}
\label{sec:analytic_model}
To derive an approximate expression for the increase in emittance caused by core depletion, the distributions of the betatron action of the incoming and outcoming bunches in a collision are calculated. To highlight the features of the process, first the simplified case with no other processes acting on the beam is studied analytically. The formulas are derived for unequal beams, and then simplified to the case of equal beams. Numerical models including other processes, such as IBS and radiation damping, are discussed in Sec.~\ref{sec:ODEs}.

The model is constructed under the approximation that the transverse distributions remain close to Gaussian during the whole store. Furthermore, it is assumed that the hourglass effect and crossing angle have a negligible influence on the shape of the distribution of the colliding particles, which makes the integrals analytically solvable. The angle and hourglass effect are instead included in an approximate way once the distribution is known. These approximations are justified in Sec.~\ref{sec:comparison}. 

The total number of particles removed, $\lumsc$, per interaction cross section $\sigtot$ during a single bunch crossing is given by an overlap integral of the densities of the two bunches~\cite{moller45,handbook98}:
\begin{equation}
\label{eq:Lsc}
   \lumsc=M N_1 N_2 \int \rho_1(x,y,s,\tau)\rho_2(x,y,s,\tau) \,\drm x\,\drm y\,\drm s\,\drm \tau,  
\end{equation}
where $M=\sqrt{(\vI-\vII)^2-(\vI\times\vII)^2/c^2}$ is a kinematic factor~\cite{moller45,furman03}. The two bunches are assumed to move with opposite velocities $\vect{v}_i$ so their centres have the longitudinal coordinate $s=\pm v\tau$. Both centres are at $s=0$ at the interaction point (IP) at time $\tau=0$. Furthermore, $\rho_i(x,y,z_i)$ is the density of bunch~$i$ normalized to one, $N_i$ its intensity, 
 and $(x,y)$ are the transverse coordinates. All integrations are to be carried out on the interval $[-\infty,+\infty]$ and this convention holds for all subsequent integrals unless indicated otherwise.

To calculate the distribution in betatron action, Eq.~(\ref{eq:Lsc}) has to be generalized to include the angular distributions of the bunches. It is assumed that the distributions in the three planes are independent and can be decoupled, which means that the crossing angle $\phi=0$ and the hourglass effect is weak. 
Then we have $\vI=-\vII$ and the transverse coordinates axes are equal for both beams. With $|\vect{v}_i|\equiv v$ the kinematic pre-factor becomes $M=2v$ and $\lumsc$ can be written as
\begin{multline}
 \label{eq:Lsc-ang-Gauss}
 \lumsc=
2v N_1 N_2 \int \rho_{1x}(x,x_1')\rho_{1y}(y,y_1')\rho_{1z}(s-v\tau)\times
\rho_{2x}(x,x_2')\rho_{2y}(y,y_2')\rho_{2z}(s+v\tau) \\ \,\drm x\,\drm y\,\drm x_1'\,\drm x_2'\,\drm y_1'\,\drm y_2'\,\drm s\,\drm \tau.
\end{multline}

The incoming bunches are assumed to be Gaussian:
\begin{eqnarray}
\label{eq:rhoGauss}
 \rho_{iu}(u,u')&=&\frac{\betSt}{2\pi \sigma_{ui}}\exp\left(-\frac{u^2+(\alphSt u+\betSt u')^2}{2\sigma_{ui}^2}\right) \\
\rho_{iz}(s\pm v\tau)&=&\frac{1}{\sqrt{2\pi}\sigma_{iz}}\exp\left(-\frac{(s\pm v\tau)^2}{2 \sigma_{iz}^2}   \right) \nonumber
\end{eqnarray}
for $u=x,y$ and $i=1,2$ for the two beams. Here $\sigma_{ui}=\sqrt{\betSt \epsilon_{ui}}$ are the transverse beam sizes, $\epsilon_{ui}$ are the transverse emittances, $\sigma_{iz}$ is the rms bunch length, and $(\betSt,\alphSt)$ are the optical parameters at the IP, which are assumed to be equal for both beams and in both planes.

Using Eq.~(\ref{eq:rhoGauss}), all integrations in Eq.~(\ref{eq:Lsc-ang-Gauss}) are now carried out, except over $x,x_1'$. The remaining integrand $\lambda$ gives the number of reactions per cross section for particles in bunch~1 in a phase space element $\drm x \,\drm x_1'$ during a single bunch crossing. Completely analogous calculations can be carried out in the vertical plane.

Normalizing by the total number of interactions, the distribution of collision points is thus
\begin{equation}
\label{eq:lambda-xxP}
 \lambda=\frac{
\betSt \sqrt{\sigma_{x1}^2+\sigma_{x2}^2} }{2 \pi \sigma_{x1}^2 \sigma_{x2} }
\exp\left({-\frac{2 x x_1' \alphSt \betSt+x_1'^2 {\betSt}^2+x^2 \left[1+{\alphSt}^2+\frac{\sigma_{x1}^2}{\sigma_{x2}^2}\right]}{2 \sigma_{x1}^2}} \right)
\end{equation}
For equal beams ($\sigma_{u1}=\sigma_{u2}$) and $\alphSt=0$, $\lambda$ simplifies to the result in Ref.~\cite{prstabBFPP09}, where it is shown that the transverse distribution of the collision points in that case is narrower than the incoming bunch by a factor $\sqrt{2}$.

Changing to action-angle variables $(J_x,\phi_x)$ through
\begin{eqnarray}
\label{eq:ac-ang}
x_1&=&\sqrt{2 J_x \betSt}\cos \phi_x \nonumber \\
x_1'&=&-\sqrt{\frac{2 J_x }{\betSt}}\left(\sin \phi_x+\alpha_{xy}^* \cos \phi_x\right),
\end{eqnarray}
and averaging over $\phi_x$, results in the distribution $\lambda_J$ of the betatron action of the colliding particles:
\begin{equation}
  \lambda_J=
\frac{{\betSt} \sqrt{\sigma_{x1}^2+\sigma_{x2}^2} I_0\left(\frac{J_x {\betSt}}{2 \sigma_{x2}^2}\right)}{\sigma_{x1}^2 \sigma_{x2}}\\
\exp\left({-\frac{1}{2} J_x {\betSt} \left[\frac{2}{\sigma_{x1}^2}+\frac{1}{\sigma_{x2}^2}\right]}\right)
\end{equation}
Here $I_0$ is a modified Bessel function.

Assuming that $m$ particles are removed in total from the bunch, the number of particles left in a small element $\drm J_x$ after the crossing is $(N_1 \rho_{J} - m \lambda_J)\drm J_x$, where $\rho_J=\exp\left(-J_x /\exI\right)/\exI$ is the \emph{incoming} distribution of $J_x$ obtained by integrating Eq.~(\ref{eq:rhoGauss}) over $\phi_x$. The density $\rhoRemain$ of the \textit{remaining} particles, normalized to unity, is thus
\begin{equation}
\label{eq:rhoRemain}
 \rhoRemain(J_x)=\frac{N_1 \rho_J-m\lambda_J}{N_1-m},
\end{equation}
The expectation value of $J_x$ of the \emph{incoming} bunch is $\exI$, while for the \emph{outcoming} bunch after the collision it is
\begin{equation}
 \tilde{\epsilon}_{1x} = \int_0^\infty J_x \, \rhoRemain(J_x) \, \drm J_x=\exI (1+\zeta)
\end{equation}
where 
\begin{equation}
 \zeta=\frac{m \sigma_{x1}^2}{2 (N_1-m) \left(\sigma_{x1}^2+\sigma_{x2}^2\right)}.
\end{equation}
Thus, the change in emittance during the crossing is $\drm \exI =\tilde{\epsilon}_{1x} -\exI=\zeta \exI$. Averaged over one turn, the emittance blowup per time becomes
\begin{equation}
\label{eq:dex/dt}
 \frac{d \exI}{dt}=\exI\zeta \frev \nip  \equiv  \frac{\exI}{\tcoll},
\end{equation}
 where $\frev$ is the revolution frequency and $\nip$ the number of IPs.

The number of removed particles is given by $m=\lumsc\sigtot$, inserting Eq.~(\ref{eq:rhoGauss}) in Eq.~(\ref{eq:Lsc-ang-Gauss}) and carrying out all integrations. So far, the luminosity reduction factor $R$, including the hourglass effect and the crossing angle, has been neglected. An approximate way of including it, which according to comparison with the tracking simulation in Sec.~\ref{sec:comparison} is accurate, is to use the distribution $\lambda_J$ calculated above but include $R$ in the calculation of $m$. We therefore have
\begin{equation}
\label{eq:m-Gauss}
 m=\sigtot R \frac{N_1 N_2 \sigtot}{2 \pi  \sqrt{(\sigma_{x1}^2+\sigma_{x2}^2) (\sigma_{y1}^2+\sigma_{y2}^2)}}.
\end{equation}
A general expression for $R$ is given in Appendix~\ref{sec:Appendix-R}. 

With Eq.~(\ref{eq:m-Gauss}), 
the rise time $\tcoll$ can be expressed in known parameters. Normally $m\ll N_i$, so for simplicity only first order in $m/N_i$ is kept. This gives
\begin{equation}
 \tcoll=\frac{4 \pi  \left(\sigma_{x1}^2+\sigma_{x2}^2\right)^{3/2} \sqrt{\sigma_{y1}^2+\sigma_{y2}^2}}{N_2 \sigtot  R \sigma_{x1}^2 \frev n_{IP}}.
\end{equation}
An analogous expression holds for the rise time of the vertical emittance. If the beams are round, the rise time of beam $i$ can be written as
\begin{equation}
\label{eq:tcoll}
 \tcoll=\frac{4 \sqrt{2} \pi \betSt \left(\exyi+\exyj\right)^{3/2}}
             {\sqrt{\exyi} N_j  \sigtot R \frev \nip },
\end{equation}
where the beam sizes have been written in terms of the transverse emittances $\exyi$. Here $i=1,2$ and $j=2$ for $i=1$ and vice versa for the two beams. If the beams are equal, we obtain finally
\begin{equation}
\label{eq:tcollEqual}
 \tcoll=\frac{16 \pi  \betSt \exy }{\nb \sigtot R \frev \nip},
\end{equation}

This is the average time rate of the emittance change over the first turn for an initially Gaussian bunch and Eq.~(\ref{eq:rhoRemain}) gives the exact non-Gaussian distribution of the outcoming bunch (neglecting crossing angle and hourglass effect). This distribution then turns in phase space during one revolution before it enters the IP again, and to obtain an exact distribution at later turns the integrations leading to Eqs.~(\ref{eq:lambda-xxP})--(\ref{eq:rhoRemain}) should be repeated recursively. 

The tracking shows, however, that for the LHC parameters the perturbation from a Gaussian is very small (see  Sec.~\ref{sec:comparison}). Therefore, Eq.~(\ref{eq:tcollEqual}) can be used also at later times without significant loss in accuracy. Another necessary condition for this approximation is that $m\ll N_1$, so that the bunch remains matched after the crossing. Otherwise the rotation in the transverse phase space has to be taken into account.

A simple model of the coupled time evolution of the bunch intensity and emittance can now be constructed. With equal beams, the instantaneous luminosity $\lum$ is~\cite{handbook98}
\begin{equation}
  \lum(t)=R(t)\frac{\nb^2(t) \kb \frev}{4\pi \exy(t)\beta_{xy}^*},
  \label{eq:lum}
\end{equation}
where 
 $\kb$ is the number of bunches.
The rate of removal of particles at an IP is given by $\sigtot \lum$, so with Eq.~(\ref{eq:lum}) we have (for \textit{one} bunch colliding at $\nip$ IPs): 
\begin{equation}
\frac{d\nb(t)}{dt}=-\sigtot R(t) \frac{\nb^2(t) \frev \nip}{4\pi \exy(t)\betSt} \equiv -\frac{\nb(t)}{T_\lum(t)}
\label{eq:nb-burnoff}
\end{equation}
Here $T_\lum$ is defined as the instantaneous lifetime due to collisions. The time evolution is given by Eq.~(\ref{eq:nb-burnoff}) coupled with Eq.~(\ref{eq:dex/dt}), which in the case of equal beams simplifies to
\begin{equation}
\label{eq:dex/dt-simpl}
 \frac{d \exy}{dt}=\frac{16 \pi \exy(t) \betSt}{ \nb(t) \sigtot R(t) \frev \nip}.
\end{equation}
It should be kept in mind that, when a strong betatron coupling is present, the emittance increase is shared between the planes but because of the analogous emittance increase in the vertical plane, the net effect for each plane is still given by Eq.~(\ref{eq:dex/dt-simpl}). 


In the general case, when a crossing angle is present, $R$ is a function of the emittances and therefore time dependent. The system of ODEs then has to be solved numerically. If there is no crossing angle, or when its effect is very small, $R$ is constant and the ODEs have the analytic solution
\begin{eqnarray}
\label{eq:exy-nb-depl}
\nb(t)&=& \frac{\nb(0)}{ \left(1+\frac{5 t}{\tcollZer} \right)^{4/5}} \nonumber \\
\exy(t)&=&\exy(0)        \left(1+\frac{5 t}{\tcollZer}\right)^{1/5}
\end{eqnarray}
where $\tcollZer=\tcoll$ at $t=0$.
This solution can be compared to the case when core depletion is not taken into account. Then Eq.~(\ref{eq:nb-burnoff}) can be solved directly, assuming a constant emittance, to yield
\begin{equation}
\label{eq:nb-no-depl}
\nb(t)=\frac{\nb(0)}{1+\frac{4 t}{\tcollZer}}.
\end{equation}
Using Eq.~(\ref{eq:lum}), the luminosity at one IP with core depletion is
\begin{equation}
\label{eq:lum-depl}
 \lum(t)=\frac{\frev \kb \nb^2(0) R}{4 \pi \exy(0) \betSt 
 \left(1+\frac{5 t}{\tcollZer}\right)^{9/5}}=
\frac{4 k_b N(0) }{\tcollZer \nip \sigtot} \frac{1}{ \left(1+\frac{5 t}{\tcollZer}\right)^{9/5}}
\end{equation}
and
\begin{equation}
\label{eq:lum-no-depl}
 \lum(t)=\frac{\frev \kb \nb^2(0) R}{4 \pi \exy(0) \betSt \left(1+\frac{4 t}{\tcollZer}\right)^2}=
\frac{4 k_b N(0) }{\tcollZer \nip \sigtot} \frac{1}{ \left(1+\frac{4 t}{\tcollZer}\right)^2}
\end{equation}
without it. 
 Finally, the integrated luminosity including core depletion is given by 
\begin{equation}
 \int_0^t \lum(\tau) \,\drm \tau=\frac{\kb \nb(0)}{\nip \sigtot} 
\left(1-\frac{1}{\left(1+\frac{5 t}{\tcollZer}\right)^{4/5}}\right),
\end{equation}
while if the core depletion is neglected, it reduces to
\begin{equation}
 \int_0^t \lum(\tau) \,\drm \tau=\frac{\kb \nb(0)}{\nip \sigtot} \cdot
\frac{1}{1+\frac{4 t}{\tcollZer}} \cdot
\frac{t}{\tcollZer}.
\end{equation}
With the parameters given in Table~\ref{tab:lhc}, we have $\tcollZer=29.9$~h for $\nip=3$. This is compared with the strength of other effects in Sec.~\ref{sec:ODEs}.

\section{Multiparticle tracking}
\label{sec:tracking}
The tracking simulation program follows two bunches containing a number of macro particles. In the simulations presented here, $5\times10^4$ particles were used to represent a nominal bunch of $7\times10^7$ \pb ions in the LHC. The 6D coordinates of the particles are updated on a turn-by-turn basis by three routines: betatron motion, synchrotron motion and collisions. All other processes are neglected in order to isolate the core depletion effect.

Betatron motion is represented by a rotation in normalized phase space by an angle given by the machine tune and chromaticity. Synchrotron motion is implemented by a change in energy and longitudinal momentum. Both routines are taken from Ref.~\cite{blaskiewicz-cool07}, where a more detailed description can be found.

To simulate the collisions, the program loops through all particles and calculates for each of them an interaction probability $P_1$ as a function of its coordinates and the distribution of the opposing bunch. A random number is then sampled to determine if an interaction takes place, in which case the particle is removed. To calculate $P_1$, the movement of a particle in bunch~1 through bunch~2 at an IP is considered. If bunch~1 contains only one particle, we have $P_1=\sigtot\lumsc$, with $\lumsc$ given by Eq.~(\ref{eq:Lsc}) without approximations. 

\begin{figure}[tb]
  \centering
  \includegraphics[width=8.5cm]{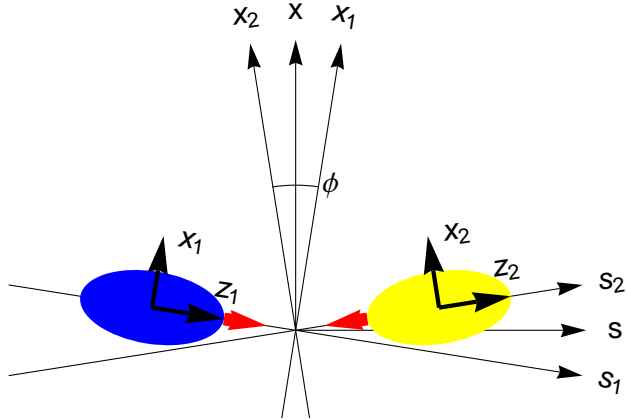}
  \caption{Schematic illustration of a collision between two bunches at an
IP. The directions of the movement are
indicated by the red arrows in the front of the bunches. The y-coordinates coincide ($y_1=y_2$). All distances are measured in the lab frame.}
  \label{fig:coord-syst}
\end{figure}

In the general case, a crossing angle $\phi\neq0$ has to be taken into account, which is assumed to be in the horizontal plane. 
The different coordinate systems used are defined in Fig.~\ref{fig:coord-syst}, where the $s_i$ axes are fixed and the $z_i$ axes move with each bunch. For simplicity, both densities will be expressed in the $s_2$--$x_2$ system. 
Since a negligible transverse magnetic field is assumed at the IP
(the experimental chambers are usually constructed in such a way that this is fulfilled)
a particle in bunch~1 
with spatial coordinates $(\xIo,\yIo,\zIo)$ at time $\tau=0$ and transverse angles $(x_1',y_1')$ 
follows approximately a straight line given by
\begin{equation}
\label{eq:x1-traj}
\begin{bmatrix} x_1 \\ y_1 \\ s_1 \end{bmatrix}=
\begin{bmatrix}\xIo-\xIo'\zIo \\ \yIo-\yIo'\zIo \\ 0\end{bmatrix}+
s_1\begin{bmatrix}\xIo' \\ \yIo' \\ 1 \end{bmatrix}
\end{equation}
where it is assumed that the longitudinal coordinate changes in time as 
\begin{equation}
\label{eq:s1(tau)}
s_1=\zIo+v\tau .
\end{equation}
In the rotated system of bunch~2, Eq.~(\ref{eq:x1-traj}) is transformed to
\begin{equation}
\label{eq:x2-traj}
\begin{bmatrix} x_2 \\ y_2 \\ s_2 \end{bmatrix}=
\begin{bmatrix} C & 0 & -S \\ 
                0 & 1 & 0         \\
                S & 0 & C \end{bmatrix}
\begin{bmatrix} x_1 \\ y_1 \\ s_1 \end{bmatrix}\equiv
\begin{bmatrix}\xIIo \\ \yIIo \\ 0 \end{bmatrix}+
s_2\begin{bmatrix}\xIIo' \\ \yIIo' \\ 1 \end{bmatrix}.
\end{equation}
where $C=\cos\phi$ and $S=\sin\phi$. Eq.~(\ref{eq:x2-traj}) contains five unknowns and three equations. The last equation can be used to express $s_1$ in terms of $s_2$, which upon insertion in the first two equations can be used to identify the remaining coefficients. The solution is
\begin{eqnarray}
 \xIIo&=&(\xIo-\xIo'\zIo)/A \nonumber\\
  \xIIo'&=&(\xIo'C-S)/A \nonumber\\
 \yIIo&=&\yIo-\yIo'(\zIo C+\xIo S)/A \nonumber\\
 \yIIo&=&\yIo'/A \nonumber\\
s_2(\tau)&=&(v \tau + \zIo) C + (\xIo + v \tau \xIo') S
\end{eqnarray}
where $A=C+\xIo'S$ and Eq.~(\ref{eq:s1(tau)}) has been used to derive the $\tau$-dependence of $s_2$.

The density function $\rho_{1}$ for a single particle, needed in Eq.~(\ref{eq:Lsc}), can be modelled by the Dirac $\delta$-function:
\begin{equation}
\label{eq:rho1-delta}
 \rho_{1}(x_2,y_2,s_2,\tau)=  
\delta(x_2-[\xIIo+\xIIo's_2])\delta(y_2-[\yIIo+\yIIo's])
\delta(s_2 - s_2(\tau)).
\end{equation}
The particles in the opposing bunch are sorted in discrete bins along the directions $(x,y,z_2)$ in order to obtain the density $\rho_2$. It is assumed that the transverse distributions are independent around the IP, and that the longitudinal density $\rho_{2z}$ does not depend on $x$ or $y$. The transverse binnings are performed using a constant $\betxy=\betSt$, but it has to be accounted for that the distribution of bunch~2 changes along $s_2$ with $\betxy(s_2)$, given by
\begin{equation}
\label{eq:beta-at-IP}
 \betxy(s_2)=\betSt\left(1+\frac{s_2^2}{{\betSt}^2}\right).
\end{equation}
This can be modelled through a parameter $\kappa(s_2)$, which gives the ratio of the width of bunch~2 at $s_2\neq 0$ to the width at $s_2=0$:
\begin{equation}
 \label{eq:kappa}
\kappa(s_2)=\sqrt{\frac{\betxy(s_2)}{\betSt}}=\sqrt{1+\frac{s_2^2}{{\betSt}^2}}.
\end{equation}
The widening of the beam is thus expressed as $\rho_{2x}(x_2,s_2)=\rho_{2x}(x_2/\kappa,0)/\kappa\equiv\rho_{2x}(x_2/\kappa)/\kappa$. 

The interaction probability $P_1=\sigtot\lumsc$ for a particle is then obtained by integrating Eq.~(\ref{eq:Lsc}) with $\rho_1$ given by Eq.~(\ref{eq:rho1-delta}). All integrations except over $s_2$ can be carried out directly. Using that the kinematic pre-factor simplifies to $M=2v\cos^2(\phi/2)$, the result is 
\begin{equation}
\label{eq:P1-coll1c}
P_1=
\frac{2 N_2 \sigtot \cos^2\frac{\phi}{2} }{d} \int
\frac{\rho_{2x}^*(\frac{\xIIo+\xIIo's_2}{\kappa(s_2)})}{\kappa(s_2)}
\frac{\rho_{2y}^*(\frac{\yIIo+\yIIo's_2}{\kappa(s_2)})}{\kappa(s_2)}
\rho_{2z}\left(\left[1+\frac{1}{A}\right]s_2-\frac{\zIo C+\xIo S}{A}\right) \,\drm s_2.
\end{equation}
Here $\rho_{2x}^*(x),\rho_{2y}^*(y)$ are the transverse densities of bunch~2 at the IP. The integral in Eq.~(\ref{eq:P1-coll1c}), which is solved on every turn for every particle, is replaced in the code by a sum over all bins that the particle passes through.

Using this mathematical model for the collisions, the core depletion effect as well as the hourglass effect are automatically accounted for. Simulation results from the tracking code are presented together with results from the analytic model in Sec.~\ref{sec:comparison}.

\section{Simulation results with collisions only}
\label{sec:comparison}

\begin{figure}[tb]
  \centering
  \includegraphics[width=8cm]{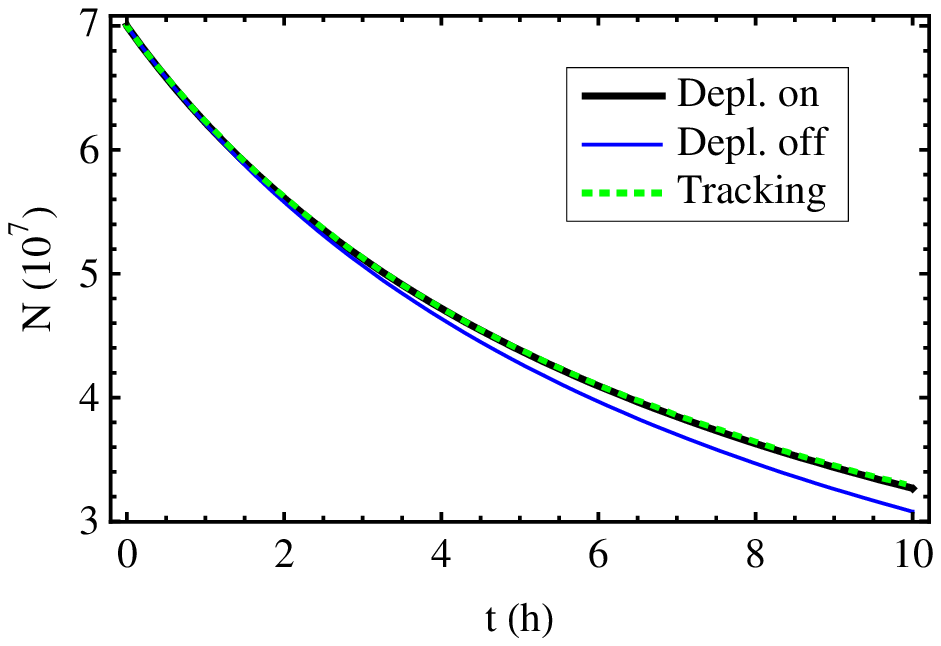}
\includegraphics[width=8cm]{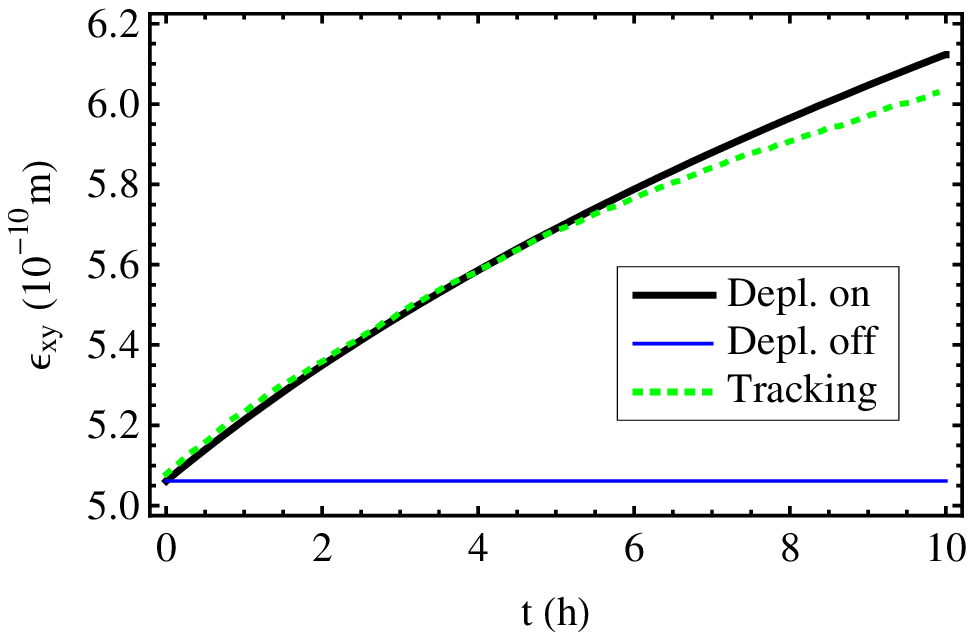}
\includegraphics[width=8cm]{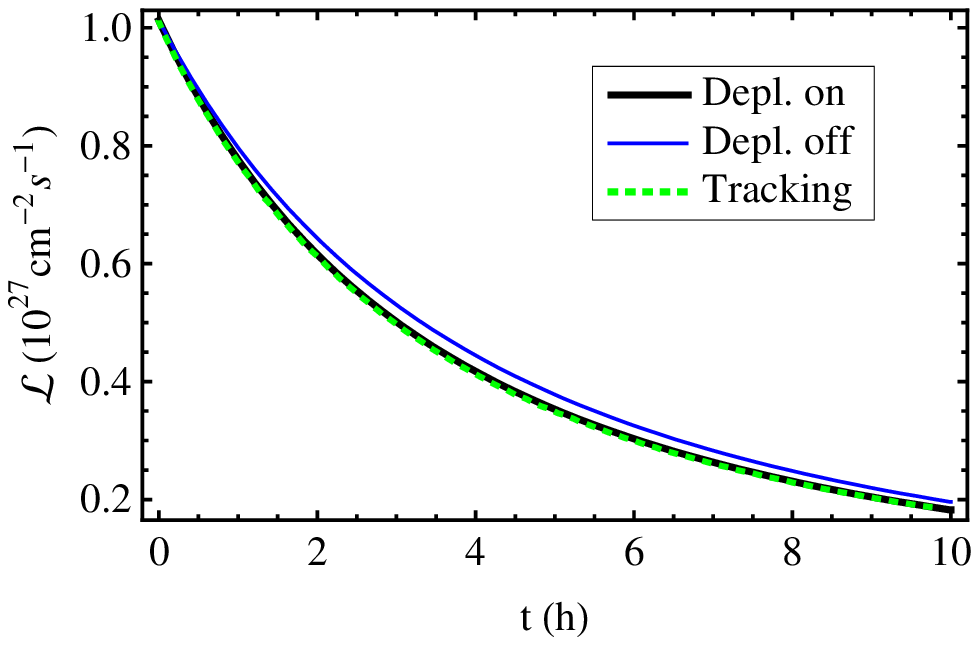}
  \caption{The time evolution of the bunch intensity $N$, the transverse emittance $\exy$ and the luminosity $\lum$ for \pb ions in the LHC, with three active IPs, when all processes but the collision are neglected, as obtained with the tracking and the analytic solution of the ODEs with and without core depletion. The crossing angle is set to $\phi=0$ at all IPs and all other parameters are given in Tab.~\ref{tab:lhc}. 
  }
  \label{fig:tracking-analytic-evol-phi0}
\end{figure}

For the comparison between the analytic model in Sec.~\ref{sec:analytic_model} and the tracking described in Sec.~\ref{sec:tracking}, two cases are considered: Either the crossing angle is $\phi=0$ at all IPs, or $\phi$ is given by Table~\ref{tab:lhc}. All other parameters are taken from Table~\ref{tab:lhc} in both cases and all three IPs are assumed active. All other processes except collisions are neglected.

The results for the first case are shown in Fig.~\ref{fig:tracking-analytic-evol-phi0}, where $\phi=0$ implies that $R$ is constant so that the analytic ODE solutions are valid. As expected, it can clearly be seen that there is an emittance increase in the tracking, which is arising solely from the variation in collision probability between the core and the tails of the beam. In this example, when other effects are not taken into account, the effective emittance increase is around 20\% over 10~h.

The agreement in bunch population and luminosity between the analytic model with core depletion included and the tracking is excellent. 
The neglection of the core depletion introduces a small error, which corresponds to a 5\% change in integrated luminosity during a 10~h store. 
A good agreement between the methods is also found for the transverse emittance, although a small difference can be seen towards the end of the store. 

\begin{figure}[tb]
  \centering
  \includegraphics[width=8cm]{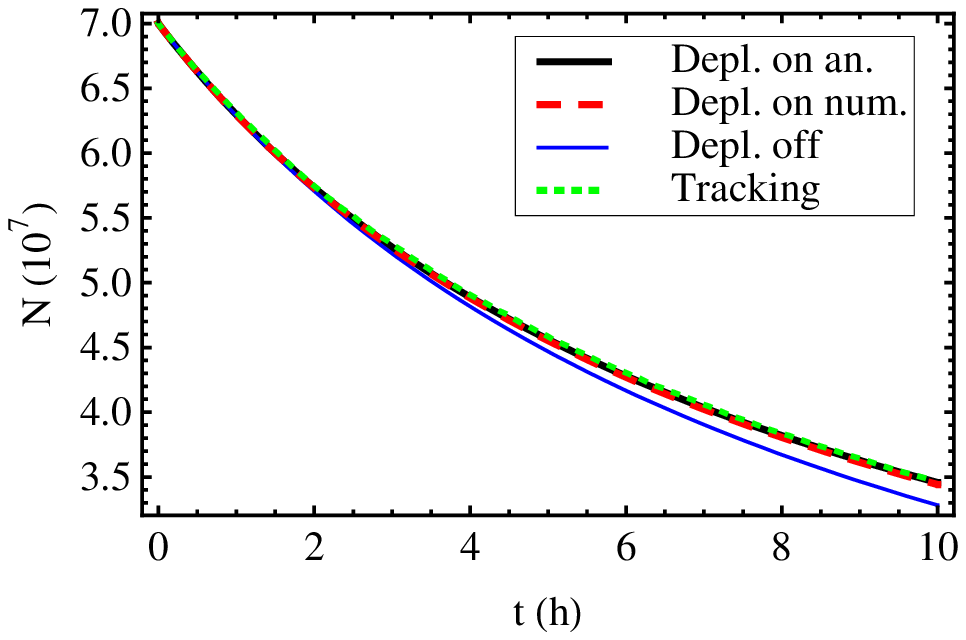}
\includegraphics[width=8cm]{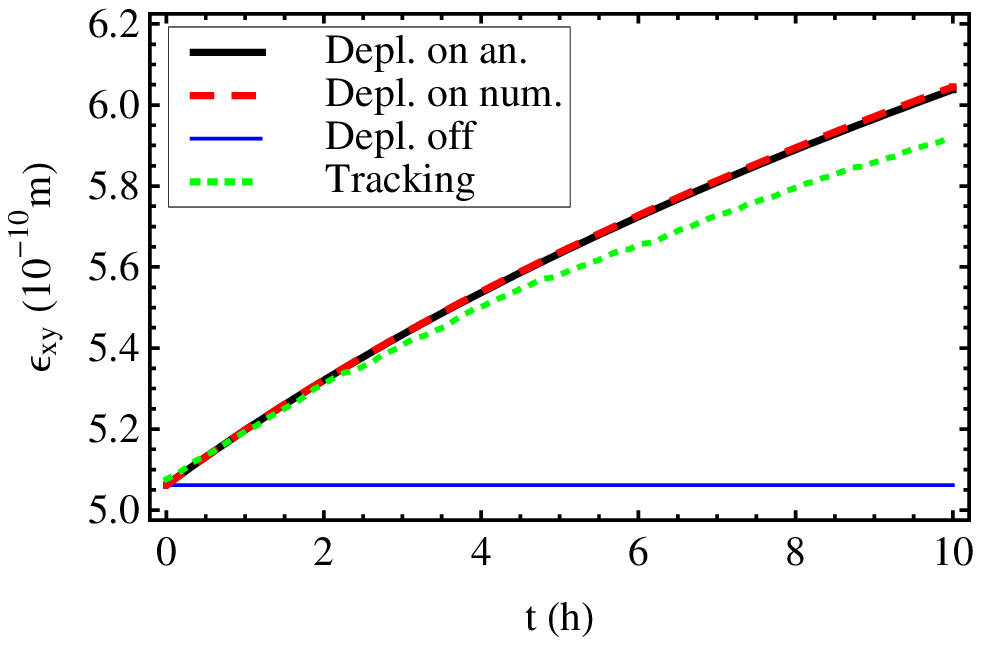}
\includegraphics[width=8cm]{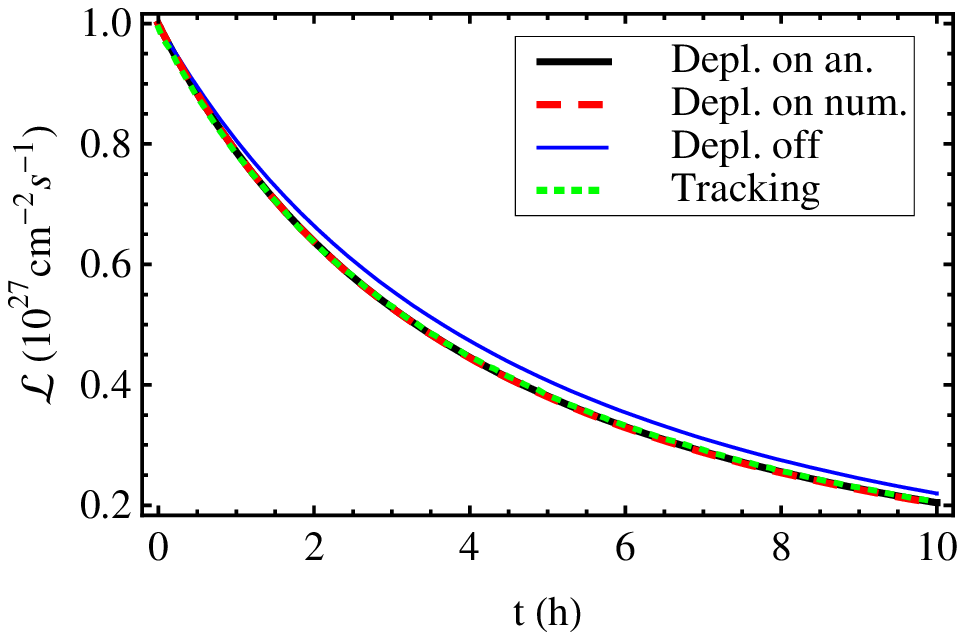}
  \caption{The time evolution of the bunch intensity $N$, the transverse emittance $\exy$ and the luminosity $\lum$ for \pb ions at IP2 in the LHC, with three active IPs, when all processes but the collision are neglected. Results obtained by the tracking simulation described in Sec.~\ref{sec:tracking}, the analytic model including core depletion (analytic and numeric solution),
  and the analytic model neglecting core depletion, given by Eqs.~(\ref{eq:nb-no-depl}) and (\ref{eq:lum-no-depl}). The numeric parameters used are given in Tab.~\ref{tab:lhc}, except the crossing angle which has been set to zero. }
  \label{fig:tracking-analytic-evol}
\end{figure}

The analogous results for the second case with a non-zero crossing angle are shown in Fig.~\ref{fig:tracking-analytic-evol}. The tracking is compared both with a numeric integration of the ODEs using Mathematica~\cite{mathematica} (with $R$ given by Eq.~(\ref{eq:rtot-gauss}) evaluated at every integration step) and the analytic solution with an assumed constant $R$ given in Table~\ref{tab:lhc}. In this case, the analytic solution is very accurate as $R$ does not change significantly. Again, an excellent agreement is found in luminosity and bunch intensity, while there is a small discrepancy in emittance. 

\begin{figure}[tb]
  \centering
  \includegraphics[width=12cm]{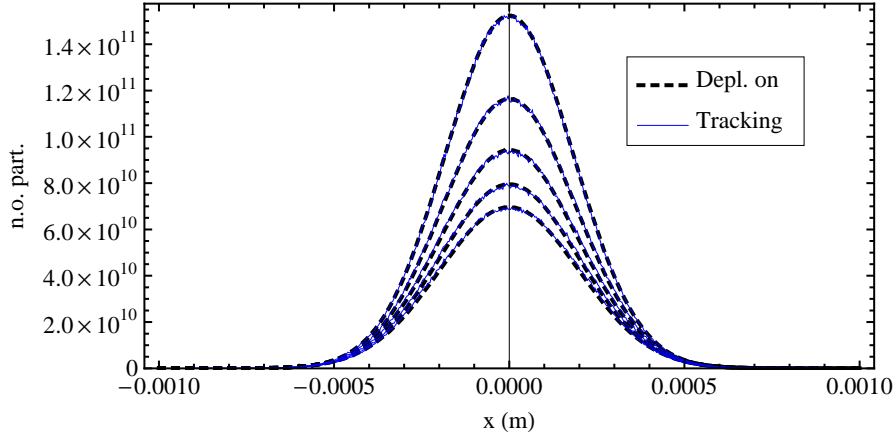}
  \caption{The transverse bunch file at different times as simulated by tracking and an analytic model with a Gaussian with the emittance given by Eq.~(\ref{eq:exy-nb-depl}). The top curve is at $t=0$ and each lower curve corresponds to a time 2.5~h later. The distributions are normalized to a $\beta$-function of 67.5~m and the integral of each curve gives the total bunch population at that time. The crossing angles are given in Tab.~\ref{tab:lhc}.}
  \label{fig:transv-profiles}
\end{figure}

A closer examination of the transverse profiles shows that the discrepancy comes from the approximation of Gaussian bunches. The distributions in the tracking are not strictly Gaussian, but very similar to a Gaussian with a larger standard deviation, which causes a small variation in emittance even though the luminosity and bunch population agree. Fig.~\ref{fig:transv-profiles} shows the bunch profile from tracking and the analytic Gaussian distribution for the case with $\phi\neq0$. 
The curves are snapshots of the distribution at 2.5~h intervals, with the uppermost curve corresponding to $t=0$~h. A small difference can be seen in the tails. 

In spite of the small discrepancy in emittance, the similarity between the two distributions throughout the store and the excellent agreement in luminosity and bunch intensity in Figs.~\ref{fig:tracking-analytic-evol-phi0} and~\ref{fig:tracking-analytic-evol} justify the approximations in the analytic model. 

\section{Luminosity time evolution including other processes}
\label{sec:ODEs}
To model a real machine, other effects such as IBS and radiation damping have to be taken into account and the longitudinal emittance $\eli$ has to be introduced as another dynamic variable. In Ref.~\cite{epac2004}, the time evolution of the bunch population, emittance and luminosity was calculated through numerical solution of a system of coupled ODEs. In this section, an analogous calculation is carried out, both with and without the inclusion of core depletion.

The time evolution of the emittances and the bunch intensity can be described by the following system of six ODEs with $i=1,2$ (expanding the result in~\cite{epac2004}):
\begin{align}
\label{eq:ODEs}
\frac{d\exyi}{dt}=&\frac{\exyi}{\tibsxy(N_i,\exyi,\eli)}   -   \frac{\exyi}{\tradxy} + \frac{\exyi}{T_\mathrm{MCS}} + \frac{\exyi}{\tcoll(N_{j},\epsilon_{xyij},\epsilon_{lij})} \nonumber\\
\frac{d \eli}{dt}=&\frac{\eli}{\tibsl(N_i,\exyi,\eli)}     -   \frac{\eli}{\tradz}\\
\frac{dN_i}{dt}=&-\frac{N_i}{T_\lum(N_{ij},\epsilon_{xyij},\epsilon_{lij})} - \frac{N_i}{T_\mathrm{gas}} 
\nonumber
\end{align}
Here the following notation has been introduced: $\tradxy,\tradz$ are the radiation damping times in the transverse and longitudinal planes, $\tibsxy,\tibsl$ are the emittance rise times due to IBS, $T_\mathrm{MCS}$ is the rise time due to multiple Coulomb scattering on rest gas, and $T_\mathrm{gas}$ is the lifetime caused by inelastic scattering on rest gas. In the LHC, quantum excitation is too weak to have an influence and is therefore neglected.
To solve Eqs.~(\ref{eq:ODEs}) numerically, 
Mathematica was used, taking into account the different crossing angles and $\betSt$ at the IPs shown in Table~\ref{tab:lhc}.

The IBS rise times were calculated with \madx, where a generalized version of the Bjorken-Mtingwa model was used~\cite{bjorken83,martini85,zimmermann05}. The evaluation of $\tibsl$ and $\tibsxy$ is done off-line on a grid of points and interpolated at run-time as in Ref.~\cite{epac2004}. Radiation damping times, as well as $T_\mathrm{MCS}$ and $T_\mathrm{gas}$, are calculated using standard formulas~\cite{handbook98}. In order to show the strengths of the different processes, Table~\ref{tab:rise-life-times} presents numerical values of the lifetimes and rise times in the beginning of the store using the starting parameters in Table~\ref{tab:lhc}. As can be seen, both processes related to the rest gas are negligible.


\begin{table}[tbh!]
\begin{center}
\label{tab:rise-life-times}
\caption{ Initial values of the rise times and lifetimes resulting from different processes, calculated using the values in Table~\ref{tab:lhc}.}
\begin{tabular}{|c|cc|cccccc|}
\hline 
        & \multicolumn{2}{c|}{Lifetimes} & \multicolumn{6}{c|}{Rise times and damping times} \\ \hline
        & $T_\lum$ & $T_\mathrm{gas}$ &$\tibsxy$ & $\tibsl$ & $\tcollZer$ & $\tradxy$ & $\tradz$ & $T_\mathrm{MCS}$ \\ \hline
$\nip=1$&  22.4    &    647.5         &  13.2    &  7.8     &    89.7     &   12.6    &   6.3    &  43598            \\ 
$\nip=3$&  7.5     &     647.5        &  13.2    &  7.8     &    29.9     &   12.6    &   6.3    &  43598                             \\ \hline
\end{tabular}
\end{center}
\end{table}

Eqs.~(\ref{eq:ODEs}) are based on the assumption that the beams remain Gaussian throughout the whole store. In particular, the expressions for $T_\lum$, $\tcoll$ and the IBS rise times are only valid for this case. It was shown in Sec.~\ref{sec:comparison} that the collisions themselves only cause small deviations from a Gaussian distribution and measurements at RHIC~\cite{fischer02} have shown similar results for IBS. Furthermore, radiation damping can be represented by a multiplication of the oscillation amplitudes by a decay coefficient, which is the same for all particles. Therefore, this does not change the shape of the distribution, only the standard deviation. Finally, the  beam-gas scattering processes are too weak to have any significant influence. Therefore, it is a fair approximation to assume that the beams keep their Gaussian shape throughout the store.

\begin{figure}[tb]
  \centering
  \includegraphics[width=7.6cm]{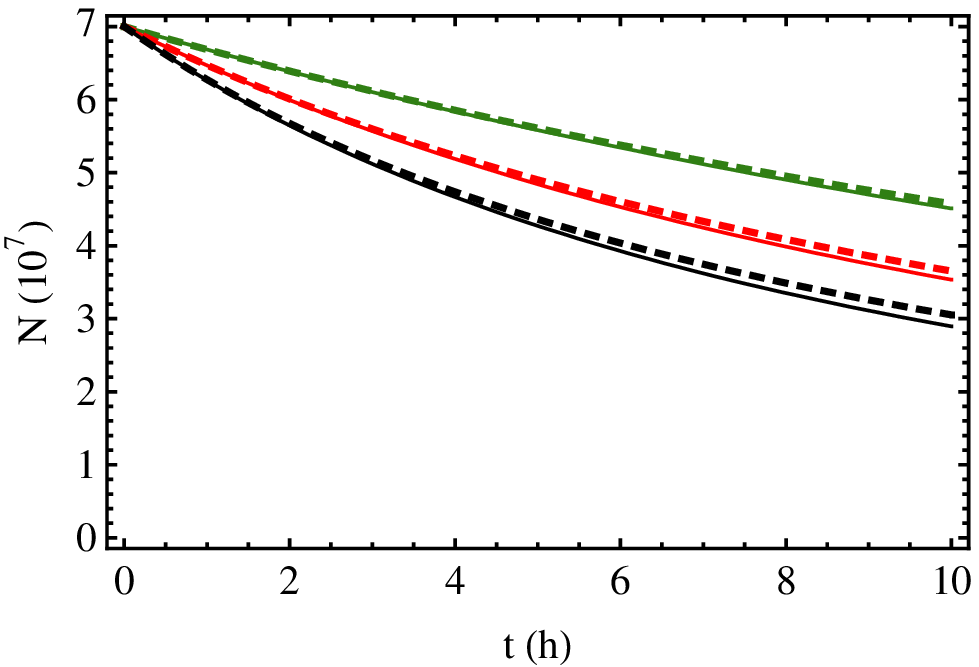}
\includegraphics[width=8cm]{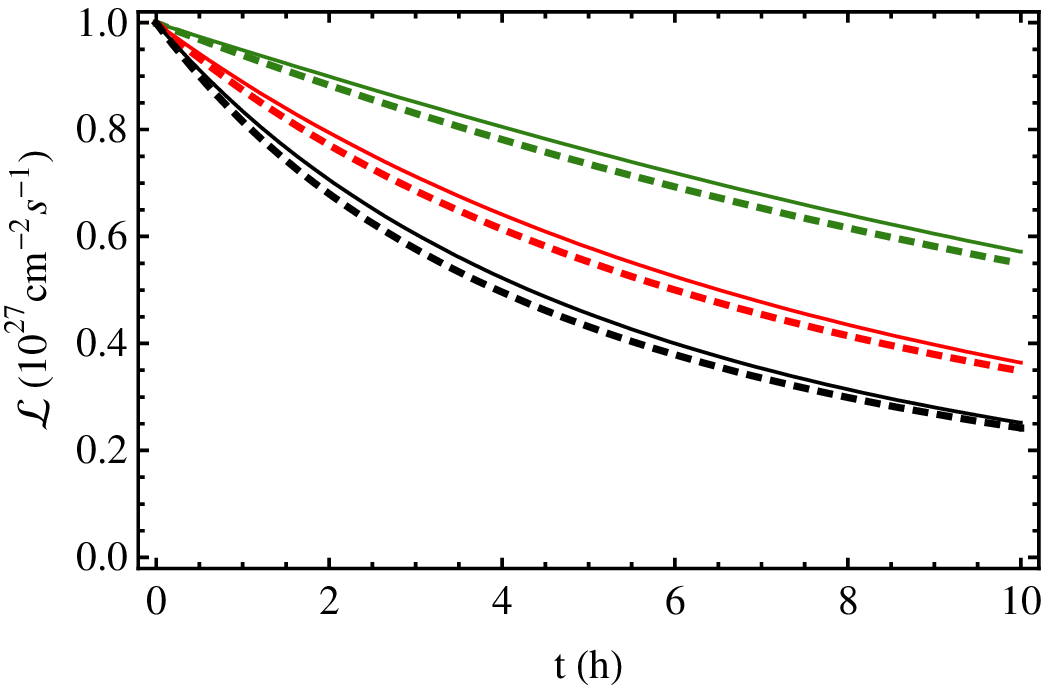}
\includegraphics[width=8cm]{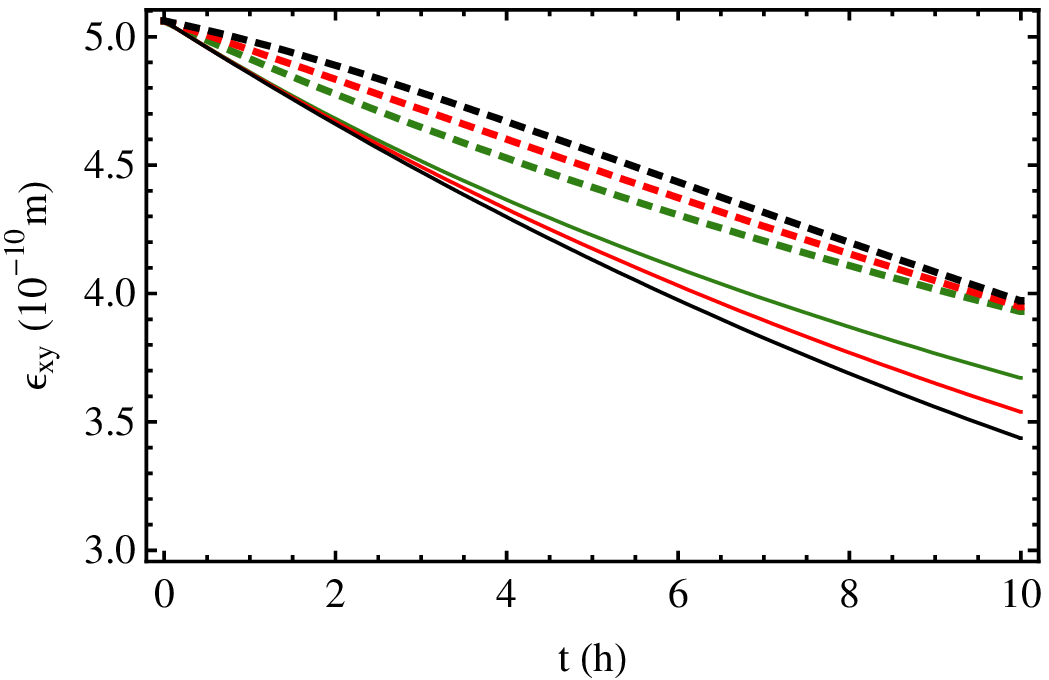}
\includegraphics[width=8cm]{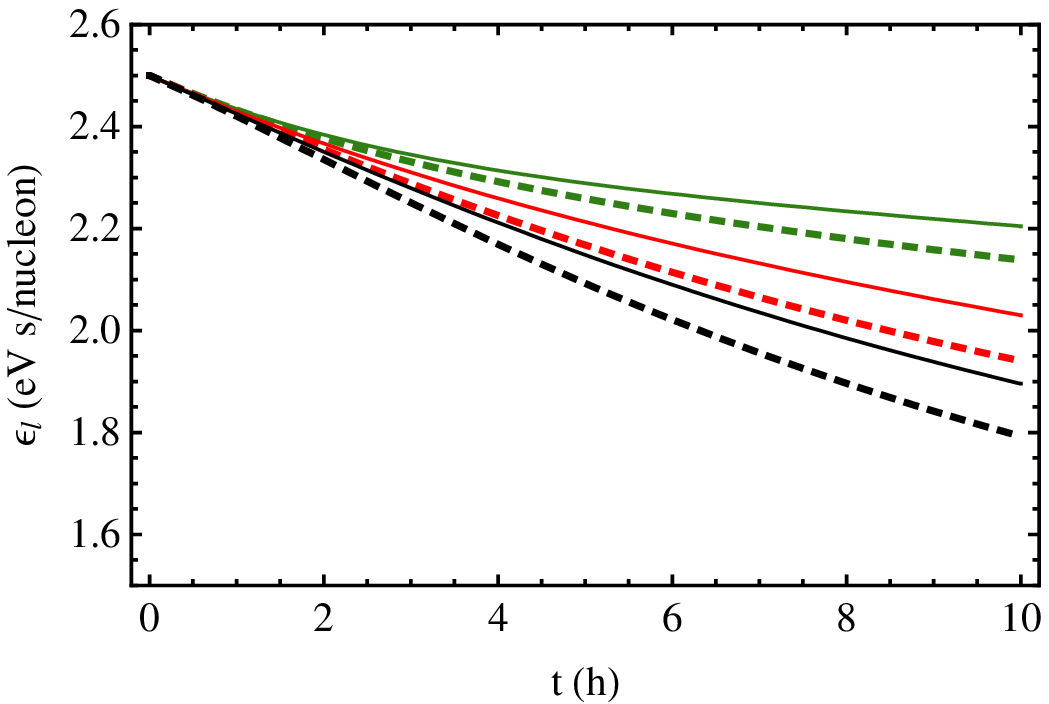}
  \caption{The time evolution, given by Eqs.~(\ref{eq:ODEs}), of the bunch intensity, luminosity at IP2, transverse rms emittance and longitudinal emittance during a 10~h store at top energy with colliding \pb beams in the LHC. Results are shown from calculations with (dotted lines) and without (solid lines) core depletion for the cases of one (green lines), two (red lines) or three (black lines) active IPs taking collisions. When only one IP is active, IP2 has been chosen. }
  \label{fig:ODE-sol}
\end{figure}

In the case of equal beams, which is studied here, the system~(\ref{eq:ODEs}) contain only three equations. The last term in the first equation represents the core depletion and to study its effect Eqs.~(\ref{eq:ODEs}) were solved also with this term excluded for $\nip=1,2,3$. The resulting luminosity, bunch intensity and emittances are shown in Fig.~\ref{fig:ODE-sol}. As can be seen, the emittance is shrinking, since radiation damping is stronger than IBS. There is a small but notable difference between the luminosity with and without core depletion. The ratio of the integrated luminosity over 10~h with core depletion included to the case without it is 0.97 for one active IP and 0.96 for three.

A striking difference can be seen in the qualitative behaviour of the time evolution of the transverse emittance in Fig.~\ref{fig:ODE-sol}. Without core depletion, the emittance shrinks faster with more IPs active, since more particles are removed through collisions in this case. Therefore the effect of IBS becomes weaker with time while radiation damping is independent of the intensity. When core depletion is included, the emittance shrinks instead faster with only one active IP, as $\tcoll$ scales linearly with $\nip^{-1}$ as shown in Eq.~(\ref{eq:tcollEqual}). This qualitative behaviour is an important observation which could be checked experimentally once the LHC is operational.

The longitudinal emittance is also shrinking, but in this case the emittance shrinks faster when core depletion is taken into account. This can be understood by considering that IBS is weaker in the longitudinal plane when the transverse emittances are larger. 


\section{Conclusions}
A new effect that increases the emittance in circular colliders with Gaussian beam profiles has been presented. Since the interaction probability in the collisions is much higher in the centre of the bunch than in the tails, the core of the beam is depleted, so that the emittance of the surviving particles is larger. 

The effect has been studied first through a simple multiparticle tracking simulation, which makes no assumptions on the shape of the beam distribution. The results show that the emittance is indeed growing due to the collisions and that the transverse distribution remains close to Gaussian. 

To describe the effect analytically, the expectation value of the betatron action was calculated before and after a bunch crossing with an initially Gaussian bunch. The emittance increase was averaged over time to form an ODE, which coupled with another ODE describing the evolution of the bunch intensity describes the time evolution under the assumption that the distribution remains Gaussian. Results from the analytic model agree very well with the tracking.

Finally a system of ODEs including also other effects changing the beam distribution and intensity, such as IBS, radiation damping and scattering on rest gas, was solved numerically for the case of \pb operation in the LHC. It was shown that a qualitatively different behaviour of the transverse emittance is expected when core depletion is included: The emittance is larger when more IPs are active, as opposed to the expected behaviour without core depletion. This prediction could be verified experimentally. Quantitatively, the core depletion effect is expected to introduce corrections of 3--4\% to the existing calculations of integrated luminosity for \pb beams in the LHC. 

\section{Acknowledgements}
I would like to thank J.M. Jowett for valuable discussions,
checking some of the calculations, and for providing his implementation of the
ODE system. 
I am also very grateful to M. Blaskiewicz for providing the core of the
tracking simulation program used here,
and to W. Fischer, S. Gilardoni, M. Giovannozzi and F. Zimmermann for helpful discussions.

\appendix
\section{Luminosity reduction factor}
\label{sec:Appendix-R}
Reductions to the luminosity due to the hourglass effect~\cite{furman91} and a non-zero crossing angle~\cite{muratori02} are well-known phenomena. For the purposes of this text, the results from Ref.~\cite{muratori02} will be rewritten on a slightly different form. 

To obtain a general reduction factor for the luminosity with Gaussian bunches, relative to the limiting case without crossing angle and hourglass effect, we start from the total number of interactions in a bunch crossing given by Eq.~(\ref{eq:Lsc}). Using the coordinate systems defined in Fig.~\ref{fig:coord-syst}, the beam distributions in the $z_i$-system moving with bunch~$i$ can be written as 
\begin{equation}
\label{eq:gaussian-dists}
\rho_i(x_i,y,z_i)=\frac{\exp\left(-\frac{x_i^2}{2 \sigma_{xi}^2}\right)}{\sqrt{2 \pi}\sigma_{xi}}
\frac{\exp\left(-\frac{y_i^2}{2 \sigma_{yi}^2}\right)}{\sqrt{2 \pi}\sigma_{yi}}
\frac{\exp\left(-\frac{z_i^2}{2 \sigma_{zi}^2}\right)}{\sqrt{2 \pi}\sigma_{zi}}.
\end{equation}
All integrations are carried out in the fixed $x-s$ system, using the transformation~\cite{muratori02}
\begin{eqnarray}
 x_i&=&x\,\cos\psi_i+s\,\sin\psi_i \nonumber \\
 y_i&=&y \nonumber \\
 s_i&=&-x\,\sin\psi_i+s\,\cos\psi_i
\end{eqnarray}
where $\psi_1=\phi/2$ and $\psi_2=-\phi/2$. Inserting the transformed distributions in Eq.~(\ref{eq:Lsc}), the $t$, $x$, and $y$-coordinates can be integrated to yield
\begin{equation}
\label{eq:lumsc-gauss}
 \lumsc=\frac{N_1 N_2}{2\pi\betSt\sqrt{(\epsilon_{x1}+\epsilon_{x2})(\epsilon_{y1}+\epsilon_{y2})}}R,
\end{equation}
where $R$ is a total reduction factor coming from the crossing angle and the hourglass effect, which, using Eq.~(\ref{eq:beta-at-IP}), is given by
\begin{equation}
\label{eq:rtot-gauss}
 R=\frac{\sqrt{2}\cos\psi_1}{\sqrt{\pi(\sigma_{z1}^2+\sigma_{z2}^2)}}
\mathlarger{\mathlarger{\mathlarger{\int}}}\frac{\exp\left(-2s^2 
\left[ \frac{\cos^2\psi_1}{\sigma_{z1}^2+\sigma_{z2}^2}  +  
\frac{\sin^2\psi_1}{\betSt(1+s^2/{\betSt}^2)(\epsilon_{x1}+\epsilon_{x1})}  \right]\right)}
{(1+s^2/{\betSt}^2)}
\drm s.
\end{equation}
In the case of equal beams, Eq.~(\ref{eq:lumsc-gauss}) is equivalent to Eq.~(2) in Ref.~\cite{muratori02}. The integral in Eq.~(\ref{eq:rtot-gauss}) is not analytically solvable in the general case and was therefore integrated numerically throughout this text. With zero crossing angle the integral simplifies to the results in Ref.~\cite{furman91} and if the hourglass effect is neglected (short bunches) the reduction factor with finite crossing angle in Ref.~\cite{muratori02} is obtained.


\begin{thebibliography}{10}

\bibitem{lhcdesignV1}
O.~S. Br{\"{u}}ning, P.~Collier, P.~Lebrun, S.~Myers, R.~Ostojic, J.~Poole, and
  {P.~Proudlock (editors)}.
\newblock {LHC} design report v.1 : The {LHC} main ring.
\newblock {\em CERN-2004-003-V1}, 2004.

\bibitem{epac2004}
J.~M. Jowett, H.~H. Braun, M.~I. Gresham, E.~Mahner, A.~N. Nicholson, and
  E.~Shaposhnikova.
\newblock {Limits to the Performance of the LHC with Ion Beams}.
\newblock {\em Proc. of the European Particle Accelerator Conf. 2004, Lucerne},
  page 578, 2004.

\bibitem{moller45}
C.~M\o{}ller.
\newblock General properties of the characteristic matrix in the theory of
  elementary particles.
\newblock {\em K. Danske Vidensk. Selsk. Mat.-Fys. Medd.}, 23(1), 1945.

\bibitem{handbook98}
{A.W. Chao, M. Tigner (editors)}.
\newblock {\em {Handbook of Accelerator Physics and Engineering}}.
\newblock World Scientific, 1998.

\bibitem{furman03}
M.~A. Furman.
\newblock The {M\o{}ller} luminosity factor.
\newblock {\em LBNL-53553, CBP Note-543}, 2003.

\bibitem{prstabBFPP09}
R.~Bruce, D.~Bocian, S.~Gilardoni, and J.~M. Jowett.
\newblock Beam losses from ultraperipheral nuclear collisions between pb ions
  in the large hadron collider and their alleviation.
\newblock {\em Phys. Rev. ST Accel. Beams}, 12(7):071002, Jul 2009.

\bibitem{blaskiewicz-cool07}
M.~Blaskiewicz and J.~M. Brennan.
\newblock Bunched beam stochastic cooling simulations and comparison with data.
\newblock {\em Proceedings of {COOL} 2007, Bad Kreuznach, Germany}, page 125,
  2007.

\bibitem{mathematica}
\url{http://www.wolfram.com}.

\bibitem{bjorken83}
J.~Bjorken and S.~Mtingwa.
\newblock {\em Particle Accelerators}, 13:115, 1983.

\bibitem{martini85}
M.~Conte and M.~Martini.
\newblock Intrabeam scattering in the cern antiproton accumulator.
\newblock {\em Part. Acc.}, 17:1, 1985.

\bibitem{zimmermann05}
\mbox{F. Zimmermann}.
\newblock {Intrabeam Scattering with Non-Ultrarelativistic Corrections and
  Vertical Dispersion for MAD-X}.
\newblock {\em CERN-AB-2006-002}, 2005.

\bibitem{fischer02}
W.~Fischer, R.~Connolly, S.~Tepikian, J.~van Zeijts, and K.~Zeno.
\newblock Intra-beam scattering measurements in {RHIC}.
\newblock {\em Proc. of the European Particle Accelerator Conf. 2002, Paris,
  France}, page 236, 2002.

\bibitem{furman91}
M.~A. Furman.
\newblock Hourglass effects for assymetric colliders.
\newblock {\em Proc. of the Particle Accelerator Conf. 1991, San Fransisco,
  California}, page 422, 1991.

\bibitem{muratori02}
B.~Muratori.
\newblock Luminosity and luminous region calculations for the {LHC}.
\newblock {\em CERN LHC Project Note 301}, 2002.

\end{thebibliography}
\end{document}